# RAPID PROTOTYPING FOR SLING DESIGN OPTIMIZATION

Zaimović-Uzunović, N.*; Lemeš, S.*; Ćurić, D.** & Topčić, A.***
*University of Zenica, Fakultetska 1, 72000 Zenica, Bosnia and Herzegovina
**Foundry "Novi život", Sarajevska 36, 72000 Zenica, Bosnia and Herzegovina
***University of Tuzla, Univerzitetska 4, 75000 Tuzla, Bosnia and Herzegovina
E-mail: nzaimovic@mf.unze.ba, slemes@mf.unze.ba, d.curic@bih.net.ba,
alan.topcic@untz.ba

**Abstract:**
This paper deals with combination of two modern engineering methods in order to optimise the shape of a representative casting product. The product being analysed is a sling, which is used to attach pulling rope in timber transportation. The first step was 3D modelling and static stress/strain analysis using CAD/CAE software NX4. The slinger shape optimization was performed using Traction method, by means of software Optishape-TS. To define constraints for shape optimization, FEA software FEMAP was used. The mould pattern with optimized 3D shape was then prepared using Fused Deposition Modelling (FDM) Rapid prototyping method. The sling mass decreased by 20%, while significantly better stress distribution was achieved, with maximum stress 3.5 times less than initial value. The future researches should use 3D scanning technology in order to provide more accurate 3D model of initial part. Results of this research can be used by toolmakers in order to engage FEA/RP technology to design and manufacture lighter products with acceptable stress distribution.

## 1. INTRODUCTION

The main objective of design optimization is to reduce manufacturing costs, thus increasing product competitiveness at the market. In each optimization method, it is necessary to choose the optimization parameter. The major optimization parameter usually refers to costs, although the cost analysis is not always achievable.

Structural analysis is used in product development to predict the product states in response to external loads. It is usually performed using finite element analysis of model generated by CAD software. In modern product development process, the use of physical prototype is minimized or even eliminated due to the ever-increasing demands for high quality, low cost products with short development time [1]. The number of design iterations can be reduced significantly if proper optimization method is implemented. However, the numerical method and structural analysis of virtual model in most cases requires the physical prototype in order to confirm the analysis results.

Rapid prototyping techniques are increasingly affordable nowadays and they can be of great help to validate numerical analysis results. These technologies shortened the development cycle.

Structural optimization is a class of optimization problems where the evaluation of an objective function(s) or constraints requires the use of structural analyses (typically FEA). It can be symbolically expressed in a compact form as [2]:

minimize f(x)
subject to g(x) ≤ 0
h(x) = 0

$$x \in D \quad (1)$$

where x is a design variable, f(x) is an objective function, g(x) and h(x) are constrains, and D is the domain of the design variable. Both g(x) and h(x) are vector functions. The design variable x is typically a vector of parameters describing the geometry of a product. For example, x, f(x), g(x), and h(x) can be product dimensions, product weight, a stress condition against yielding, and constraints on product dimensions, respectively. Depending on the definition of design variable x, its domain D can be continuous, discrete or the mixture of both. Also, a variant of structural optimization has multiple objectives, where the objective function is a vector function f(x), rather than a scalar function f(x) [1].

Variations of structural optimization expressed in the form of equation (1) can be roughly classified into geometry parameterization (size, shape or topology), approximation methods, optimization algorithms, and the integration with non-structural issues [1]. In sizing parameterization, design variable x is a predefined set of the dimensions that describe product geometry. Sizing optimization is typically done in conjunction with feature-based variational geometry [3].

A review of shape optimization based on the direct geometry manipulation approaches can be found in [4], where the boundary representations are classified as polynomials, splines, and design elements.

As a hybrid of the direct and indirect geometry manipulation approaches, the Traction Method was proposed [5, 6], where the boundary sensitivities are replaced with the velocities of boundary changes in response to the fictitious loads that deform the original shape to the target shape defined by the boundary sensitivities. By this replacement, sufficient smoothness of the boundary can be achieved via the direct manipulation of the product boundary. Optimization methods used recently include Nonlinear Programming Algorithms, Reliability and Robustness Optimization Methods and other methods.

## 2. PROBLEM STATEMENT

### 2.1 Structural optimization

Structural design optimization of 3D CAD model of a sling was performed through following phases:
- Defining initial design including 3D modelling, setting elements for static analysis, choosing material, finite element meshing, initial static analysis;
- Choosing manufacturing technology, choosing material, defining optimization boundary conditions;
- Optimization of design shape in order to minimize total slinger weight (objective function) whilst keeping equivalent Von-Mises stress at initial level (constraint function);
- According to optimization results, estimating the final design which corresponds to adopted manufacturing technology, producing model and prototype of slinger. Performing redesigning including static analysis;
- Rapid prototyping of the new product;
- Preparing mould from RP patterns;
- Testing new product characteristics.

### 2.2 Methods and techniques used

Structural design optimization of slinger is performed using following methods and techniques:
- 3D modelling and static analysis was performed using CAD/CAE software NX4;

- Slinger shape optimization was performed using Traction method, by means of software Optishape-TS;
- To define constraints for shape optimization; i.e. to exclude certain areas out of optimization domain, FEA software FEMAP was used.

**2.3 Description of product being optimized**

The slings consist of steel rope and other elements, and they are being used for various purposes: industrial use, timber processing, construction works, mostly for heavy-weight lifting (Figure 1). The rope ends are pressed within aluminium or steel shells in order to protect the rope wires. Depending on the purpose, slings can be equipped with additional accessories, such as hooks, rings, sliders etc.. All commercially used slings have to be tested according to appropriate standards.

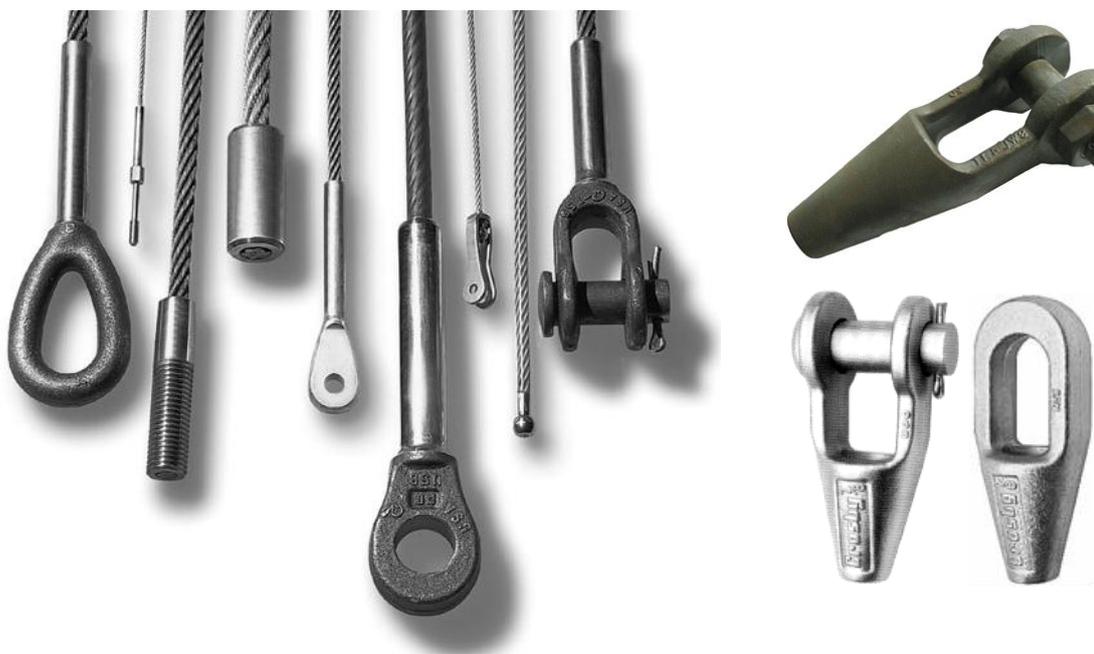

Figure 1: Various sling designs.

## 3. INITIAL SLING DESIGN

For initial static analysis, the 3D model presented in Figure 2 was used. The 3D model was built according to available technical specifications (2D drawings).

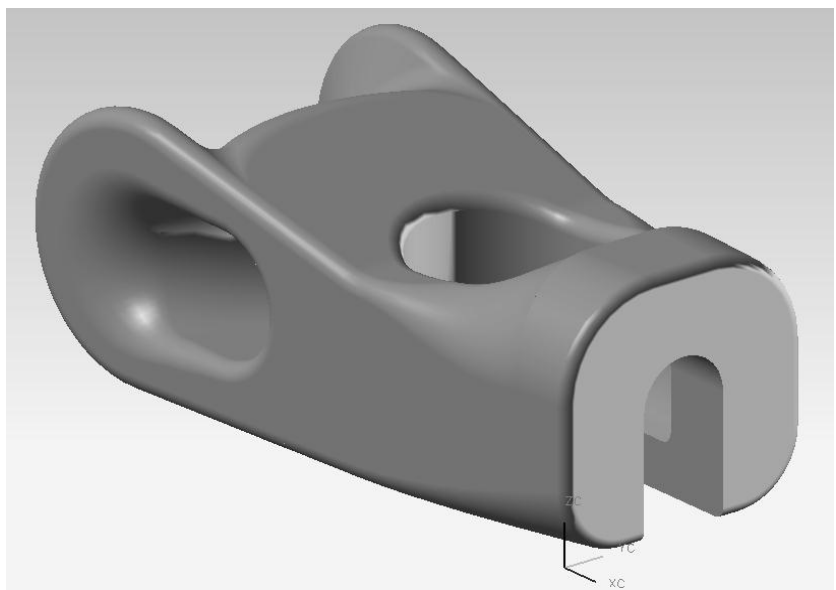

Figure 2: 3D model of initial design.

The static analysis of initial sling design was performed with the following material characteristics:

- specific mass 7850 kg/m$^3$
- Young's modulus 2E+11 N/m$^2$
- Poisson ratio 0.25
- Plasticity limit 225 N/mm$^2$
- Yield point 440 N/mm2
- Allowed tension stress 150 N/mm$^2$

### 3.1 Static analysis

Initial static analysis was performed with 110 kN force acting perpendicular to curved opening surface, and with pin constraint at the shell supporting point. The results are shown in Figure 3.

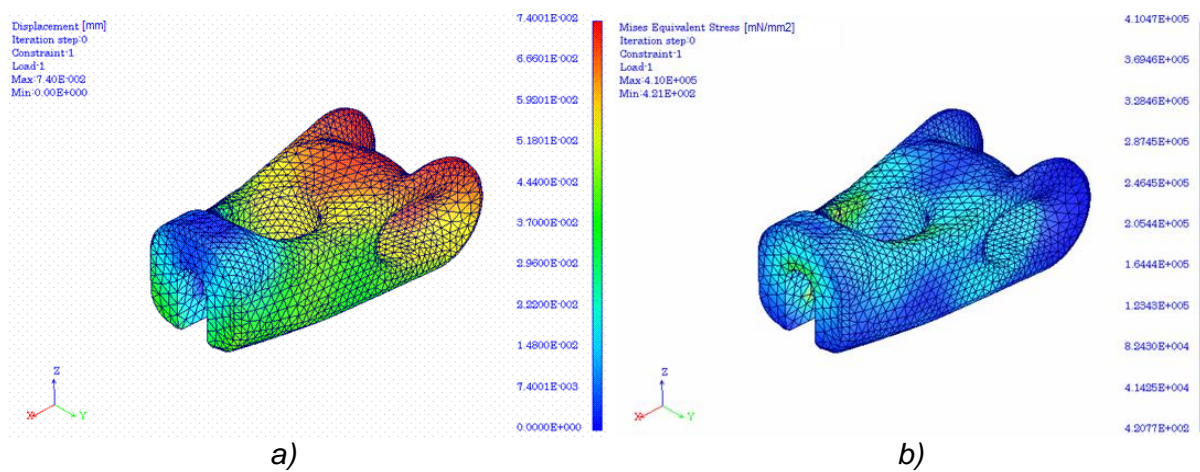

*a)* *b)*

Figure 3: Initial static analysis results: a) displacements, b) equivalent stress.

274

## 4. SHAPE OPTIMIZATION

Figure 4 shows the initial 3D model with optimization constraints. The highlighted lines represent areas which are excluded from shape optimization process in order to maintain the function (holding and fixating the steel rope).

### 4.1 Optimization results

After 30 steps of optimization performed using software Optishape-TS, the graph was plotted, as shown in Figure 5.

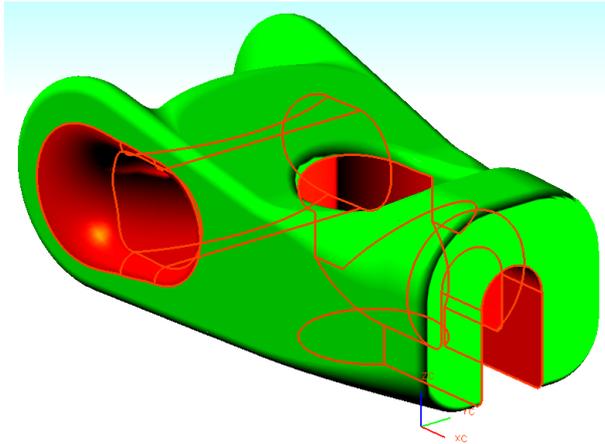 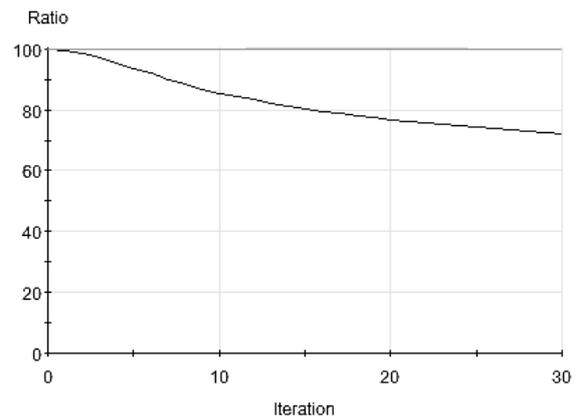

Figure 4: Optimization constraints.   Figure 5: Optimization graph.

The 28% mass reduction of initial mass was achieved, but it had to be decreased onto 20%, due to manufacturing technology limitations. Figure 6 shows the initial design and Figure 7 shows the optimized shape.

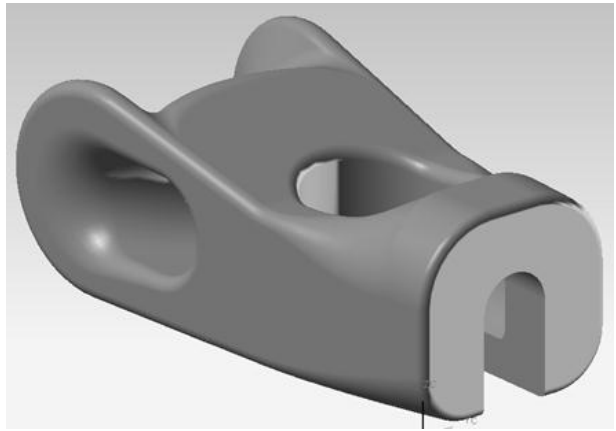 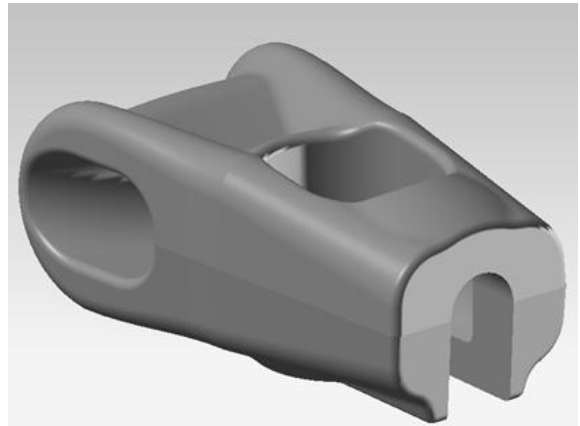

Figure 6: Initial sling design.   Figure 7: Optimized shape.

### 4.2. Final static analysis

Figure 8 shows the equivalent stress distribution in redesigned sling. When these results are compared with initial design results (shown in Figure 3), it is obvious that stresses are distributed smoothly across the whole sling body.

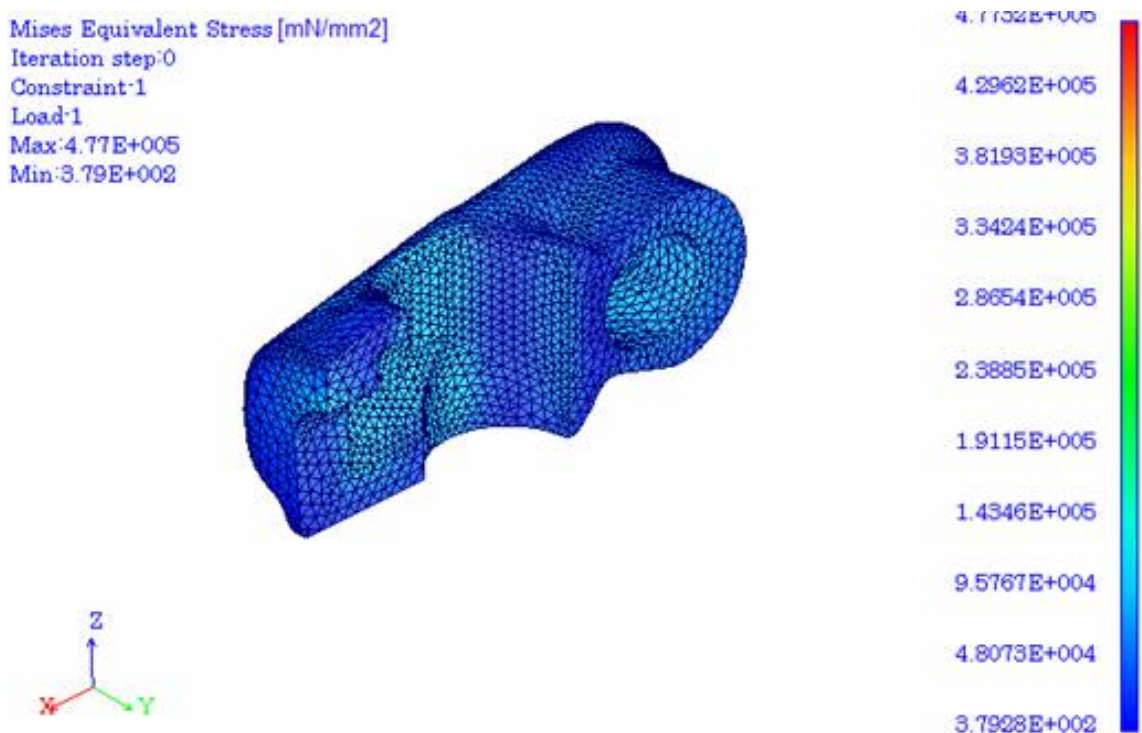

Figure 8: Equivalent stress distribution in redesigned sling.

Allowed stress for this sling is only 22 kN (3.5 times less than initial value). Therefore, the maximum equivalent stress is around 125 N/mm$^2$ which is significantly lower than allowable stress for chosen material. The equivalent stress, in most areas of sling being redesigned, is significantly lower than allowable stress. There are some peak values of stresses in the areas next to ring-shaped area where steel rope is in contact with the shell. It happens due to stress concentration because of small curvature radius and FEM mesh quality.

Figure 9 shows the analysis results, illustrating the deformation due to tension.

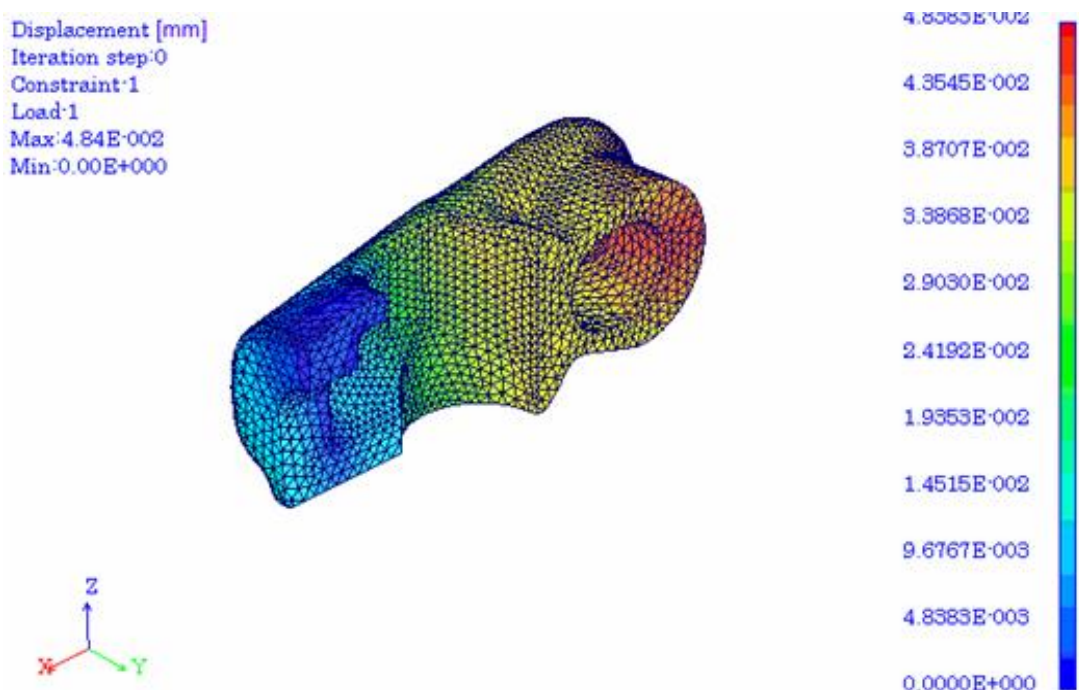

Figure 9: Deformation of sling under tension.

## 5. RAPID PROTOTYPING

3D models for rapid prototyping were prepared according to rules for casting technology. It was necessary to adjust the angles, and to predict casting channels. The 3D CAD models created for upper and lower part of redesigned sling model are presented in Figures 10 and 11. Figure 12 shows additional models necessary to manufacture the mould.

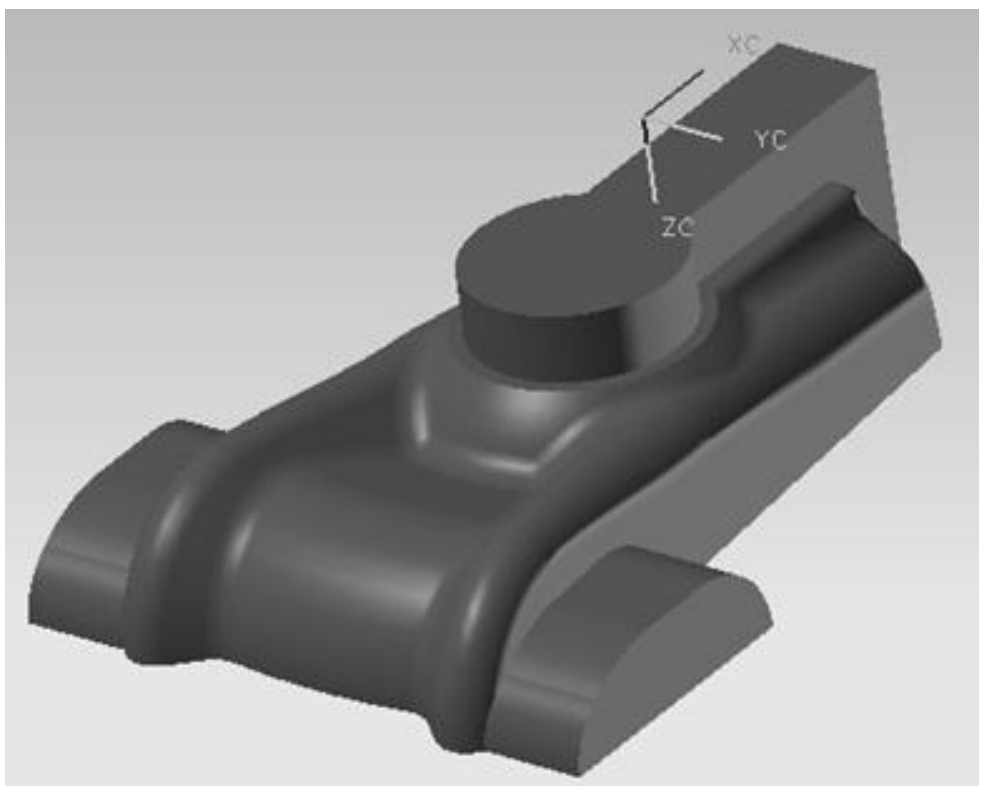

Figure 10: CAD model for lower part of redesigned sling.

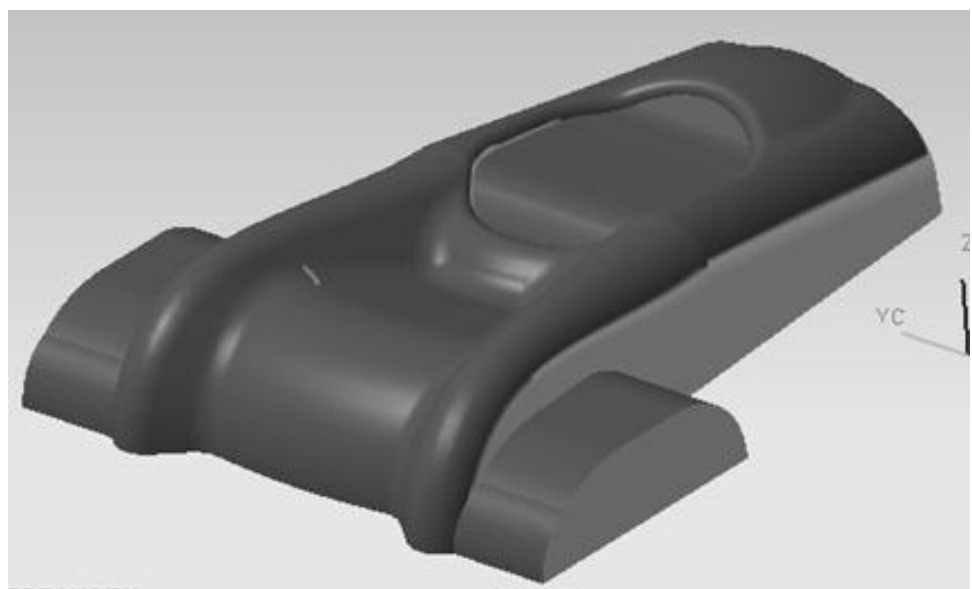

Figure 11: CAD model for upper part of redesigned sling.

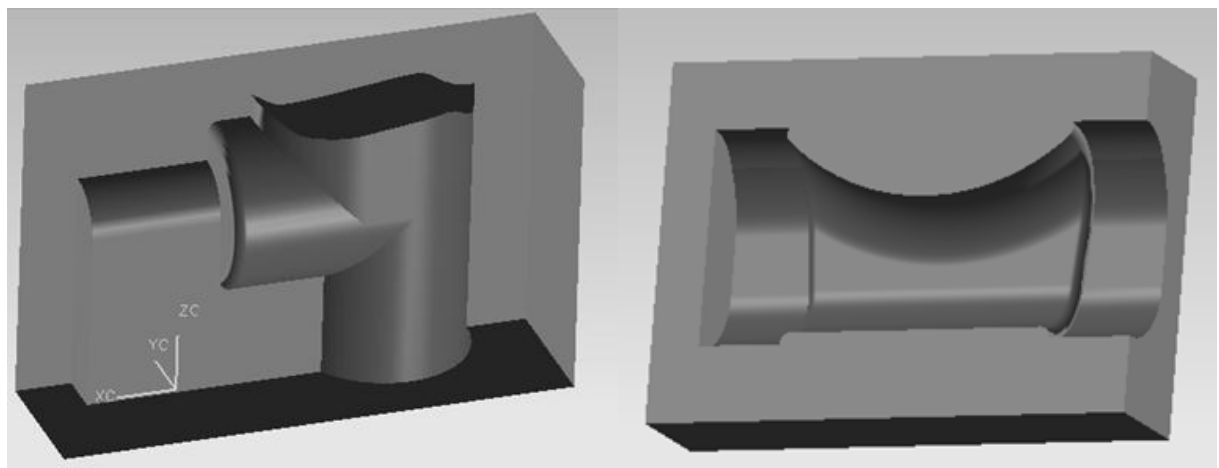

Figure 12: Other models necessary to manufacture the mould.

The 3D CAD models shown in Figures 10 to 12 were transformed to STL files and they were used in rapid prototyping machine. The Fused Deposition Modelling (FDM) method was used to build a model. The build material is thermoplastics. Figures 13 and 14 show prototypes created after shape optimisation performed in this research.

The following step is to manufacture the tool for casting the real model, which will later be tested in order to confirm the shape optimization results. The rapid prototype will be used as starting shape for the mould.

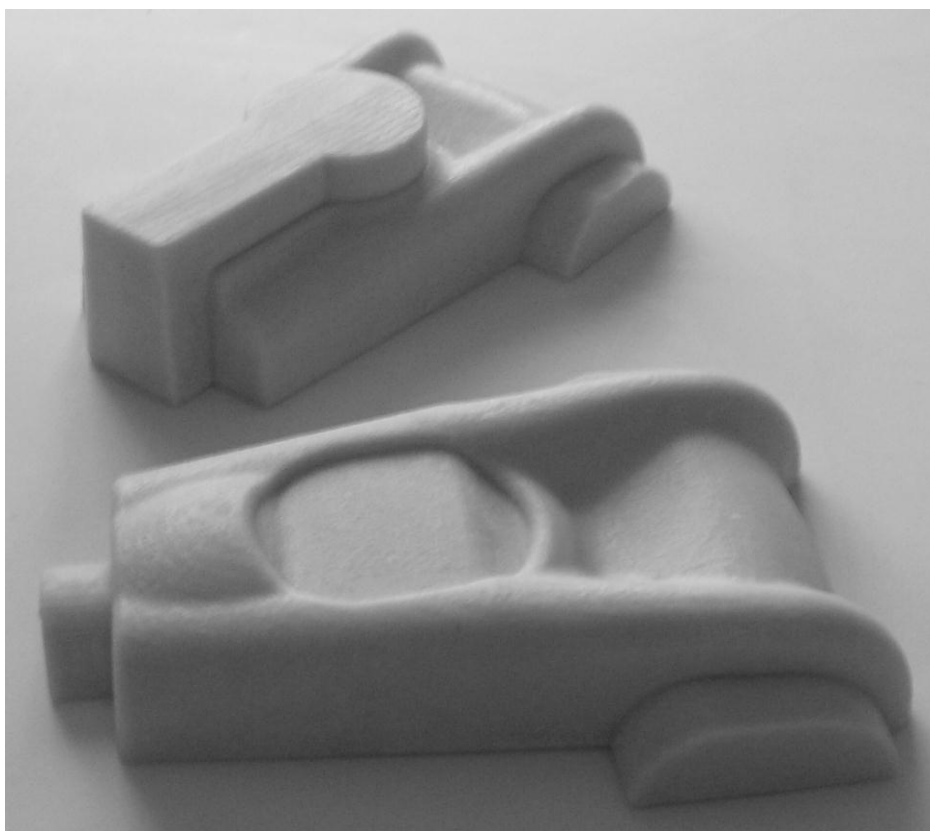

Figure 13: Rapid prototypes of upper and lower part.

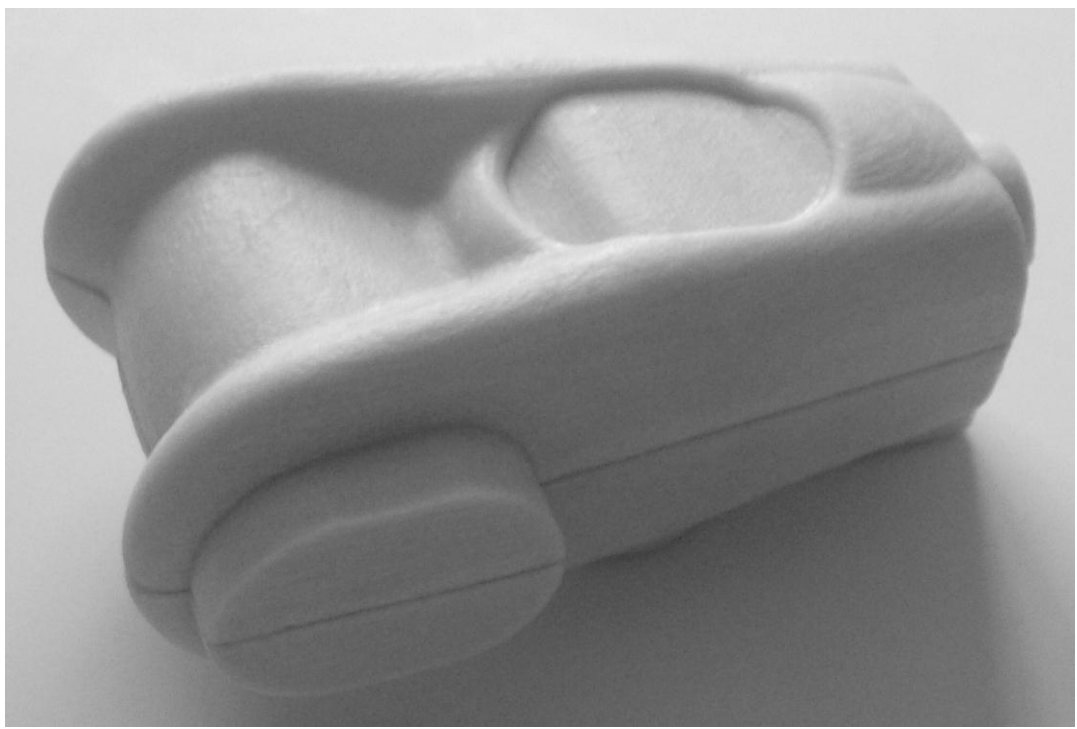

Figure 14: Connected parts of rapid prototype.

## 6. CONCLUSION

The concept of design and optimization presented here can be used in wide range of design optimization cases, where it is not possible to use methods based on design parameterization, due to irregular and complex configuration. These products do not consist of regular simple geometric features.

To solve this problem, a number of commercial CAE software solutions were developed. This research used set of software packages (NX-4, Optishape-TS, Femap) to perform design shape optimization. These software packages were used to obtain the global optimum (optimum solution which will obey constrains). The software Optishape-TS also offers a library of possible shapes which include shape parameters with possibility to exclude some parts of the model from optimization process. They are usually those parts/areas which will not change its shape during optimization.

This method shortens the development cycle, and gives an opportunity to reduce the mass or other cost-related parameter.

During this optimization process, it is possible to define more optimization objectives (objective functions) and constraint functions. It is also possible to exclude some areas of the shape being optimized from optimization process. In this case, dimensions of front and back openings (used for the steel rope connection with this sling) had fixed values.

The goal of this research was to reduce the mass and this goal was realized; the slinger mass is reduced by 20-30%, even obeying the rules for casting technology (angles, wall, thickness, casting channels,...).

The rapid prototyping techniques are now affordable methods for fast product development. The engineering practice uses a lot of computational methods to achieve the design which will fulfil all demands. Anyway, no numerical method is 100% accurate, and it is necessary to manufacture the physical model to be tested before the product is manufactured in larger series.

The rapid prototyping techniques, such as Fused Deposition Modelling (FDM) method, create the physical model of product which can be used for manufacturing moulds or other tools.

This research was followed by casting the real products. Their performance should be tested experimentally, in order to confirm the shape optimization results.